\documentclass[usenatbib]{mn2e}

\usepackage{graphicx}
\usepackage{amssymb}
\usepackage{amsmath}
\usepackage{color}
\newcommand{\beq}{\begin{equation}}
\newcommand{\eeq}{\end{equation}}

\def\psr{{PSR~B1259--63}}

\title[The X-ray-object from \psr{}]{The origin of the X-ray-emitting object moving away from \psr{}}
\author[Barkov \& Bosch-Ramon ]{Maxim V. Barkov$^{1}$\thanks{E-mail: maxim.barkov@riken.jp (MVB)} and
Valent\'i Bosch-Ramon$^2$
 \\
$^{1}$ Astrophysical Big Bang Laboratory, RIKEN, 351-0198 Saitama, Japan \\
$^{2}$ Departament d'Astronomia i Meteorologia, Institut de Ci\`ences del Cosmos (ICCUB),\\
Universitat de Barcelona (IEEC-UB), Mart\'{\i} i Franqu\`es 1,
E-08028 Barcelona, Spain}

\begin{document}
\date{Received/Accepted}
\maketitle

\begin{abstract} 
A mysterious X-ray-emitting object has been detected moving away from the high-mass gamma-ray binary \psr{}, 
which contains a non-accreting pulsar and a Be star { whose} winds collide forming a complex interaction structure. 
Given the strong eccentricity of this binary, the interaction structure should { be strongly anisotropic, which}
together with the complex evolution of the shocked winds, could explain {the origin of} the observed moving X-ray feature. 
We propose here that a fast outflow made of a pulsar-stellar wind mixture is always present moving away from 
the binary in the apastron direction, with the injection of stellar wind occurring { at} orbital phases close 
to periastron passage. This outflow periodically loaded with stellar wind would move with a { high} speed, and likely 
host non-thermal activity due to shocks, on scales similar to those of the observed moving X-ray object. 
Such an outflow is thus a very good candidate to explain this X-ray feature. This, if confirmed, would imply 
pulsar-to-stellar wind thrust ratios of $\sim 0.1$, and the presence of a jet-like structure on the {larger}
scales, up to its termination in the ISM.
\end{abstract}
                                                                                          
                                                                                          
\section{Introduction}
\label{introduction}

High-mass gamma-ray binaries are formed by an OB-type star and a compact object, a neutron star or a black hole. The system \psr{} is presently the only one hosting a non-accreting pulsar, with a period of 47~ms and spin-down power  $L_{\rm sd} = 8.3\times10^{35}$~erg~s$^{-1}$ \citep{joh92}, and has a fast rotating Oe-type star as a companion, with mass $M\approx 30 M_{\sun}$ and luminosity $L_*=2.3\times10^{38}$~erg~s$^{-1}$ \citep{neg11}.

Recent {\it Chandra} observations of \psr{} have shown, on scales of few arcseconds, an object moving away from the system with a rather high spatial velocity of $\sim 0.1\,c$ \citep{phk15}. The data suggest that the object would have been ejected from the binary around periastron passage, and may be accelerating, although this acceleration is not statistically significant.

Here we propose a hypothesis for the origin of the X-ray emitting object moving away from \psr{} \citep{phk15} 
as resulting from an eccentric orbit and the complex two-wind interaction. First, in Sect.~\ref{model}, 
we present some analytical estimates of the main properties of the proposed scenario, and then, in Sect.~\ref{simu}, 
numerical simulations are used to further support our proposal. Section~\ref{disc} summarizes this work and provides with some additional discussion.

\begin{figure}
\includegraphics[width=85mm]{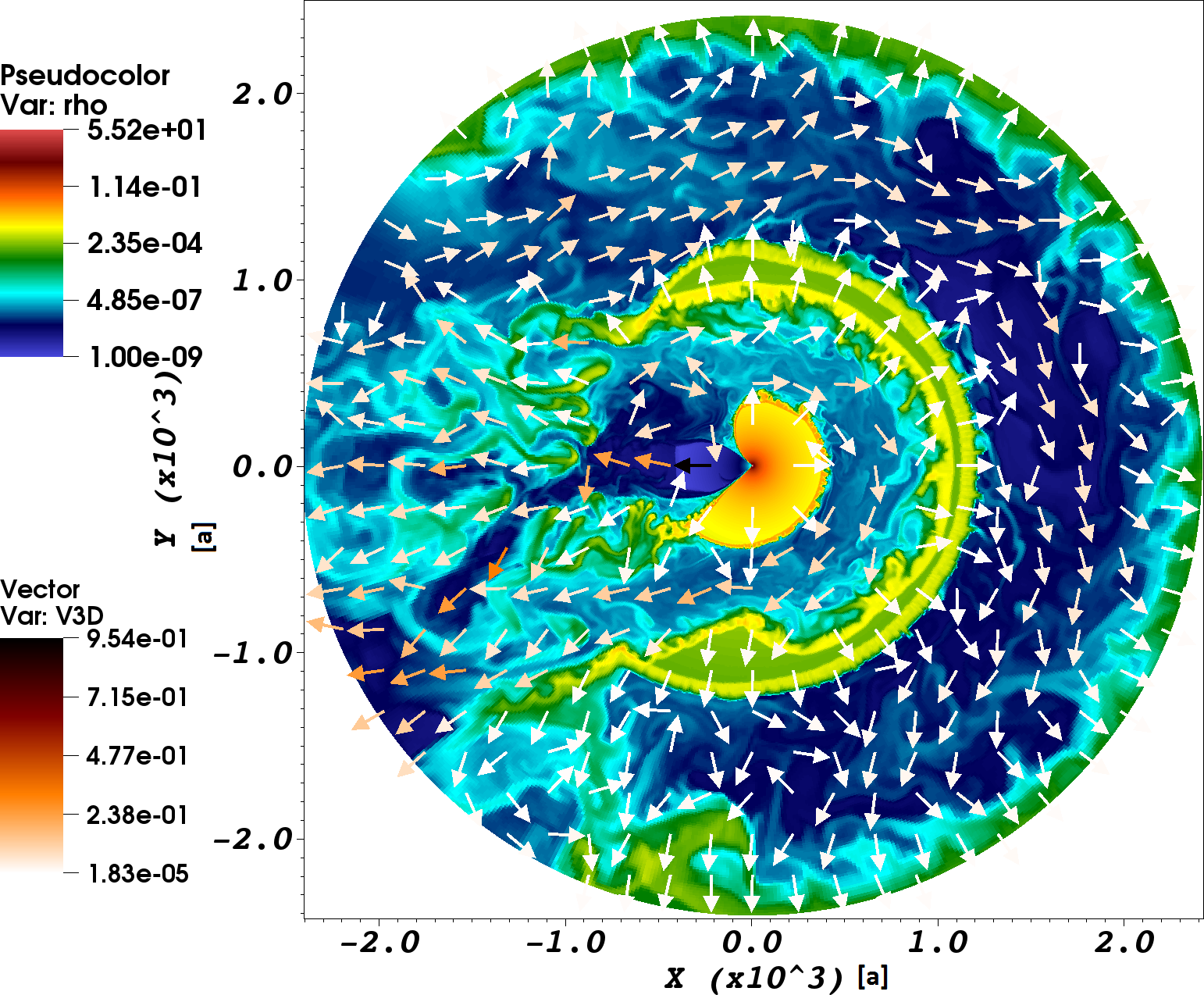}
\caption{Density distribution by colour and arrows representing the flow motion direction (in speed of light units), 
in the orbital plane ($XY$), for \psr{} after a simulation time of $t=2500$~days (680 days after periastron passage).
{ To the right from the binary, there are two semi-annular regions of shocked pulsar 
wind material, injected during $t_{\rm pe}$, and separated by shocked stellar wind. To the left, the unshocked pulsar wind freely propagates radially until is shocked due to orbital bending of 
the shocked winds, but the shocked pulsar wind keeps moving without much deflection. Not far beyond, two-wind mixing becomes apparent as a fast moving, low-density flow filled with 
slow stellar wind fragments.}}
\label{fig:psrrho}
\end{figure}

\section{The X-ray-emitting object ejected from \psr{}}
\label{model}


After the stellar and the pulsar wind collide, a shocked stellar-wind structure forms pointing in the direction of the source of the lowest-thrust wind, typically the pulsar one\footnote{Note however that energetically it is expected that the pulsar wind will strongly dominate the stellar wind \citep[e.g.][]{bbkp12}.}. This shocked-wind structure is forced to bend by orbital motion and the impact of Coriolis forces, and further out, the shocked stellar wind is radially accelerated by the shocked pulsar wind due to pressure gradients, as both shocked winds spiral away from the binary suffering instabilities and mixing \citep{bb11}. 

In a very eccentric orbit as that of \psr{}, the stellar wind injected in the apastron direction, when the pulsar is close to periastron and its wind cone points elsewhere, suffers the strongest acceleration. This occurs because the pulsar-wind cone points in that direction for most of the orbit, as the orbital velocity is very slow far from periastron. This also implies that, when pointing in the apastron direction, the shocked two-wind structure will be roughly axisymmetric \citep[as the case simulated by][]{bog08,bog12}, with bending occurring at large distances from the binary. Such a configuration, together with instabilities that disrupt the coherent shocked two-wind structures and mix the two winds, can lead to the formation of fast clumps or filaments of fragmented, shocked stellar wind material, radially moving away from the system in the apastron direction.

The pulsar-to-stellar wind thrust ratio is obtained from:
\beq
\eta=\frac{L_{\rm sd}}{\dot{M} v_{\rm w} c}\,{\rm ,}
\label{eta}
\eeq
where $\dot{M}$ and $v_{\rm w}$ are the stellar mass-loss rate and wind speed, respectively. Following \cite{eu93,bog08}, the half-opening angle of the contact discontinuity of the cone-like structure can be estimated from:  
\beq
\alpha = \frac{\pi}{6}(4-\eta^{2/5})\eta^{1/3}\,,
\label{alfa}
\eeq
{which is valid if $\eta<1$.}
The mass and speed of the stellar wind shell injected in the apastron side, while the pulsar-wind cone points elsewhere, can be estimated from the relative solid angle of the cone-like structure $\Omega$. If $\eta\ll 1$, $\Omega$ is simply $\sim \alpha^2/4$, and the injected stellar wind mass injected roughly in the apastron direction will be
\beq
M=\dot{M}\Omega t_{\rm pe}\,{\rm ,}
\label{mpw}
\eeq
where $t_{\rm pe}$ is the periastron timescale, or more properly, the pulsar residence time far from apastron. 

To obtain $t_{\rm pe}$, one requires the true anomaly ($u$) from which the stellar wind starts to be injected, at $u\sim\pi+\alpha$, as the pulsar leaves the cone-like region, until stellar wind injection stops again when the pulsar returns, at $u\sim\pi-\alpha$. Based on $u$, we can derive the eccentric anomaly ($E_{\rm a}$) from:
\beq
E_{\rm a}=2\arctan\left(\left(\frac{1-e}{1+e}\right)^{1/2}\tan\left(\frac{u}{2}\right)\right)\,{\rm ,}
\label{ecca}
\eeq
where $e$ is the orbit eccentricity, and then the mean anomaly ($M_{\rm a}$) from:
\beq
M_{\rm a}=E_{\rm a}-e \sin(E_{\rm a})\,.
\label{meana}
\eeq
The periastron time can be then expressed as
\beq
t_{\rm pe} =t_{\rm orb}\frac{\Delta M_{\rm a}}{2\pi}=t_{\rm orb}\frac{M_{\rm a}}{\pi}\,{\rm ,}
\label{tpp}
\eeq
where $t_{\rm orb}$ is the orbital period.
Assuming full pulsar and stellar wind mixing within the cone, the maximal termination speed can be estimated as
\beq
v_{\rm t} = \sqrt{\frac{2 L_{sd}(t_{\rm orb}-t_{\rm pe})}{\dot{M}\Omega t_{\rm pe}}} = \sqrt{ \frac{\eta v_{\rm w} c}{\Omega} \frac{\pi-M_{\rm a}}{M_{\rm a}}}\,.
\label{}
\eeq

\begin{figure}
\includegraphics[width=80mm]{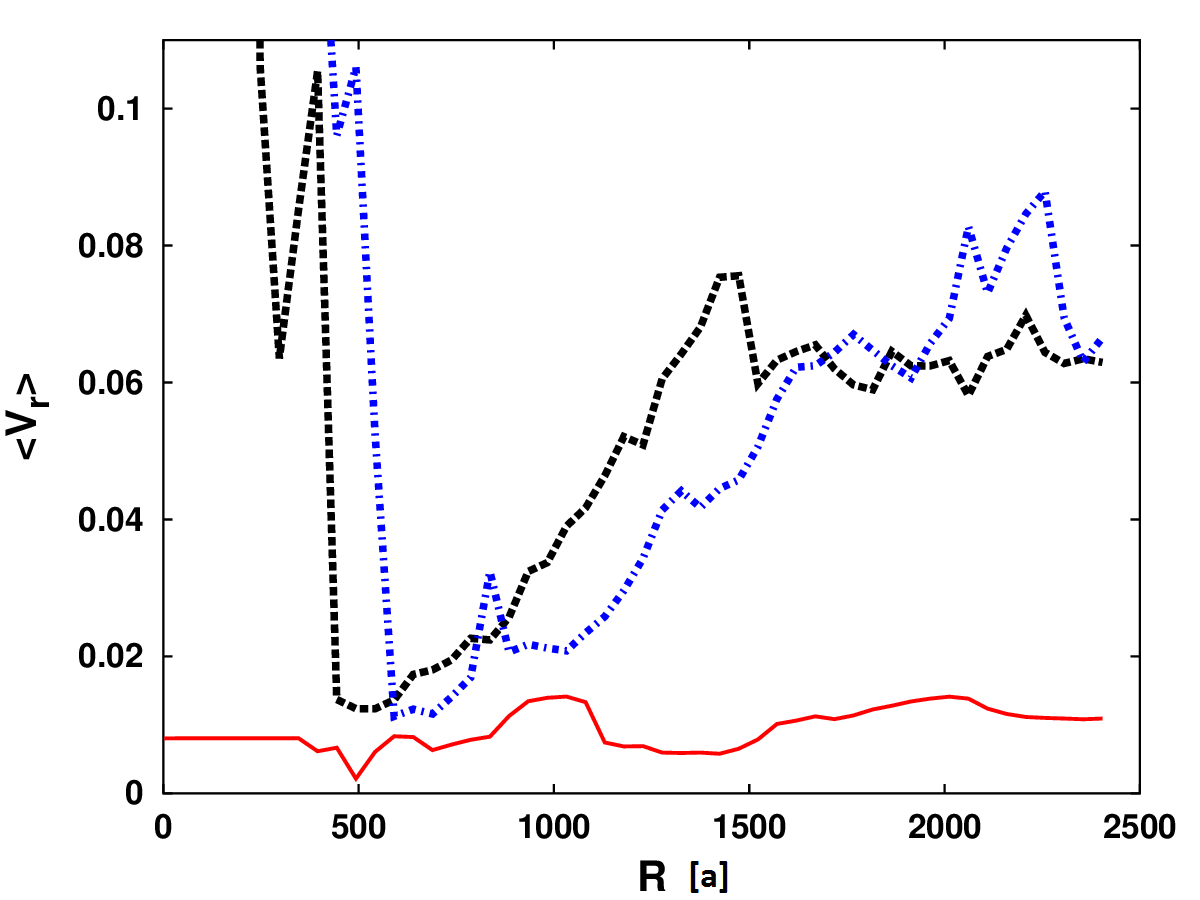}
\caption{Profile of the $\phi$-averaded radial velocity ($<V_r>$; see Eq.~\ref{vavr}) with distance from the binary for sector Sp and $t=2514$~days (692~days after periastron) (red solid line), and for sector Sa and $t=2389$~days (567 after periastron passage) (black-dashed line) and $t=2514$~days (692 days after periastron passage) (blue dot-dashed line).
}
\label{fig:psrvav}
\end{figure}

In \psr{}, the stellar wind thrust is not well known, but we can use typical late O-type star values, $\eta\sim 0.1$, {which implies $\dot{M}=1.8\times 10^{-8}\,M_\odot$~yr$^{-1}$,} and 
{ $v_{\rm w}=2400$~km/s. These values,} together with the system eccentricity, $e=0.87$ \citep{joh92}, allow the derivation of all the important geometrical parameters: $\alpha = 0.84$, $u = 2.3$, $E_{\rm a}=1.06$, $M_{\rm a} = 0.3$, yielding $t_{\rm pe} = 0.1 t_{\rm orb}$.
The mass of the stellar wind in the cone will be therefore $M\sim 5\times10^{24}$~g, and the terminal velocity of the mixed wind within the cone-like channel $v_{\rm t}\sim 0.2~c$, which is similar, but exceeds, the projected velocity $0.07\pm0.01$~c found by \cite{phk15}.  

In this scenario, the most plausible radiation mechanism behind the X-ray object moving away from \psr{} is synchrotron emission. 
The shocked pulsar wind is already supersonic far from the binary in the apastron side, and much faster than the fragments of stellar wind material. 
This leads to the formation of shocks, to an enhancement of the local magnetic field ($B$), and likely to particle acceleration. 
The expected X-ray luminosity can be related to the observed one as $L_{\rm x}\sim\chi_{\rm x}L_{\rm sd}\sim 10^{31}$~erg~s$^{-1}$ \citep{phk15}, 
so $\chi_{\rm x}\sim 10^{-6}$. The X-ray efficiency can be expressed as 
$\chi_{\rm x}\sim\chi_{\rm NT}\times\min[1,t_{\rm dyn}/t_{\rm sy}]$, where $\chi_{\rm NT}$ is the available energy fraction going to accelerate electrons, 
$t_{\rm dyn}$ the emitter dynamical timescale ($\sim 10^3$~days; \citealt{phk15}), and $t_{\rm sy}$ the synchrotron timescale. Around 1~keV, setting $\chi_{\rm NT}=0.1$,
$\chi_{\rm x}\sim 0.1\times\min[1,t_{\rm dyn}/t_{\rm sy}]\sim 10^{-6}$ can be achieved if: (i) $B$ is $\gtrsim 3\times 10^{-7}$, approximately 300 times below the magnetic 
field in equipartition with the pulsar wind energy density at $10^{17}$~cm; and (ii) the acceleration energy gain per second is $\gtrsim 10^{-3}\,q\,B\,c$, 
which is a moderate value.

\section{Simulating the wind collision in \psr{} on large scales}\label{simu}

The simulations were performed using a simplified 3 dimensional (3D) geometry in spherical coordinates using the {\it PLUTO} code\footnote{Link http://plutocode.ph.unito.it/index.html} \citep{mbm07}. Spatial parabolic interpolation, a 3rd order Runge-Kutta approximation in time, and an HLLC Riemann solver were used \citep{mig05}. {\it PLUTO} is a modular Godunov-type code entirely written in C and intended mainly for astrophysical applications and high Mach number flows in multiple spatial dimensions. The simulations were run through the MPI library in the CFCA cluster of the National Astronomical Observatory of Japan and the Great Wave FX100 Fujitsu cluster in RIKEN.
\begin{figure*}
\includegraphics[width=80mm]{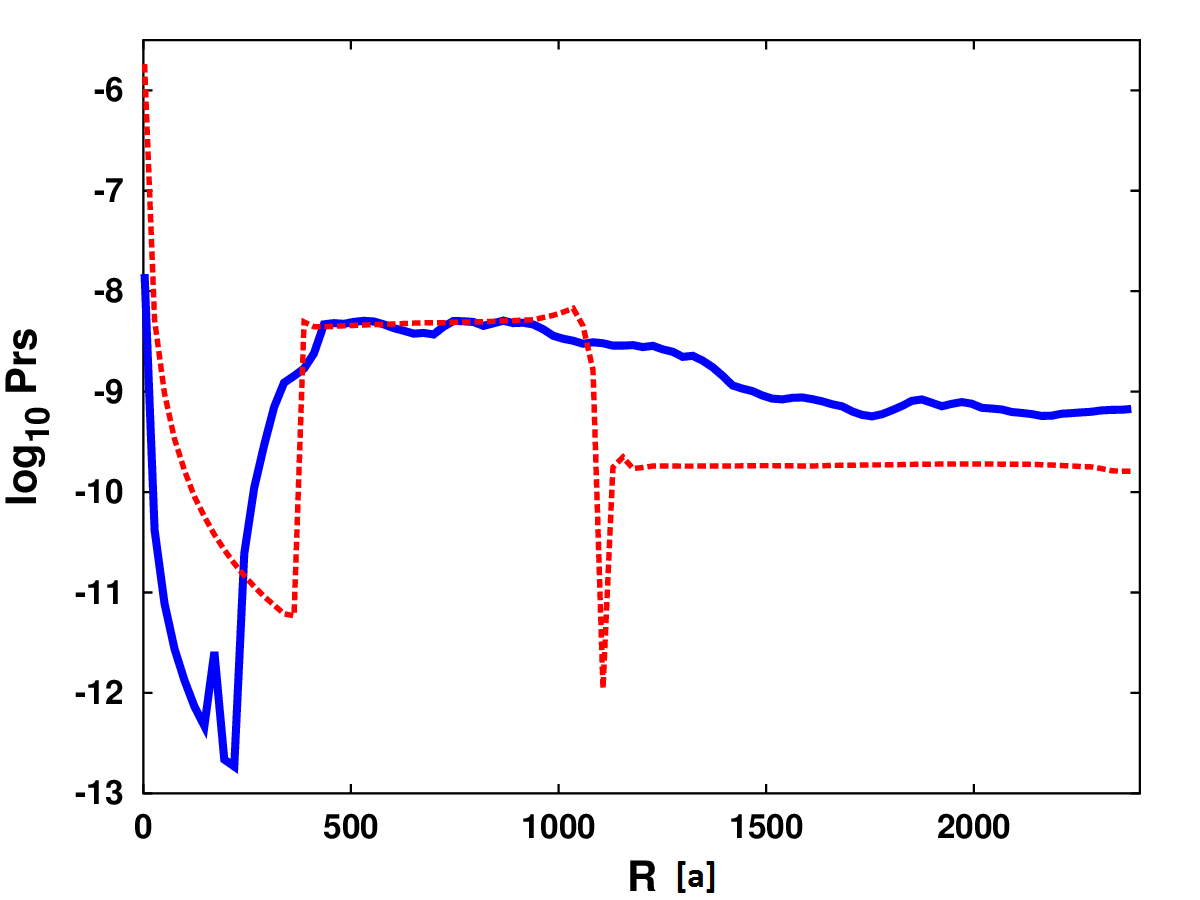}
\includegraphics[width=80mm]{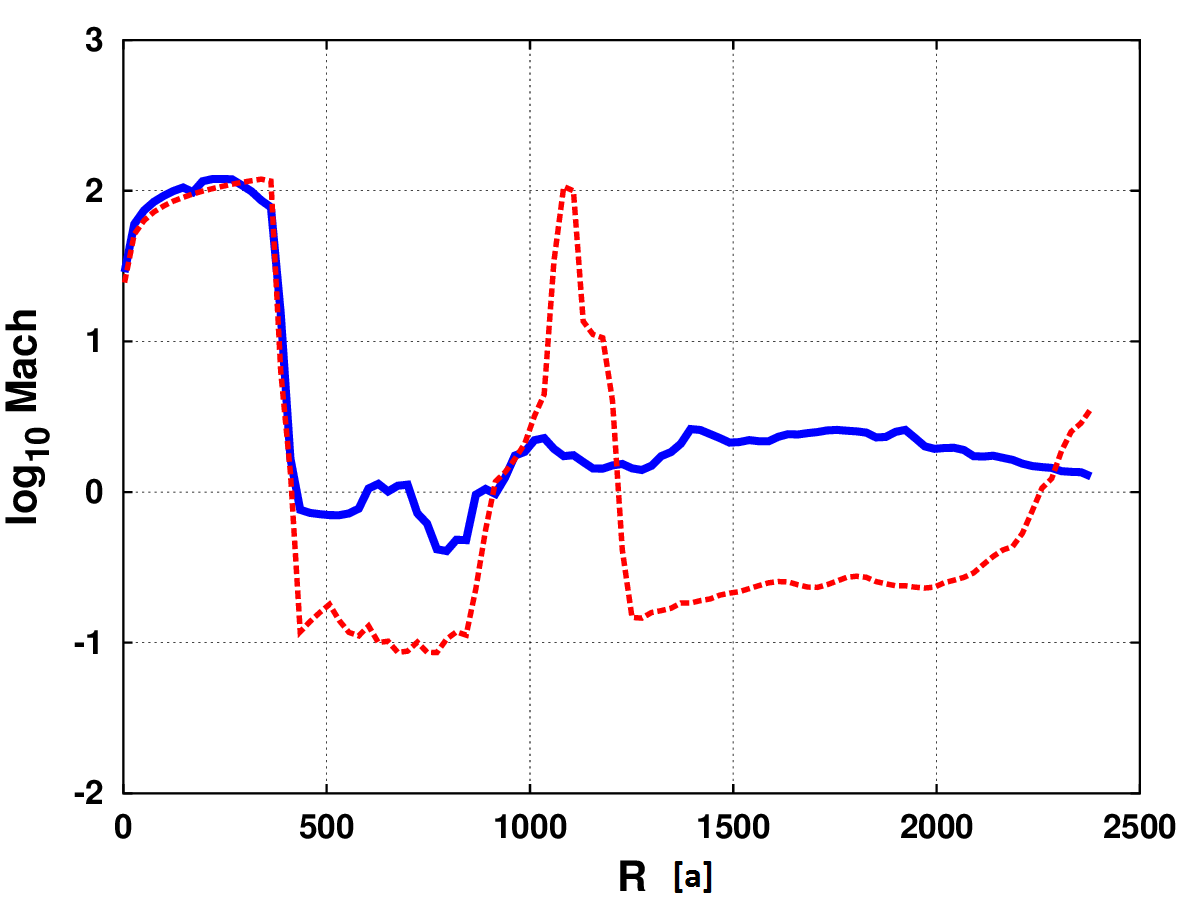}
\caption{Profile of the pressure (left panel) and Mach number (right panel) averaged by volume for sectors Sa (blue solid line) and Sp (red dashed line), and $t=2500$~days (680 days after periastron passage).}
\label{fig:psrrpcm}
\end{figure*}

\begin{figure}
\includegraphics[width=88mm]{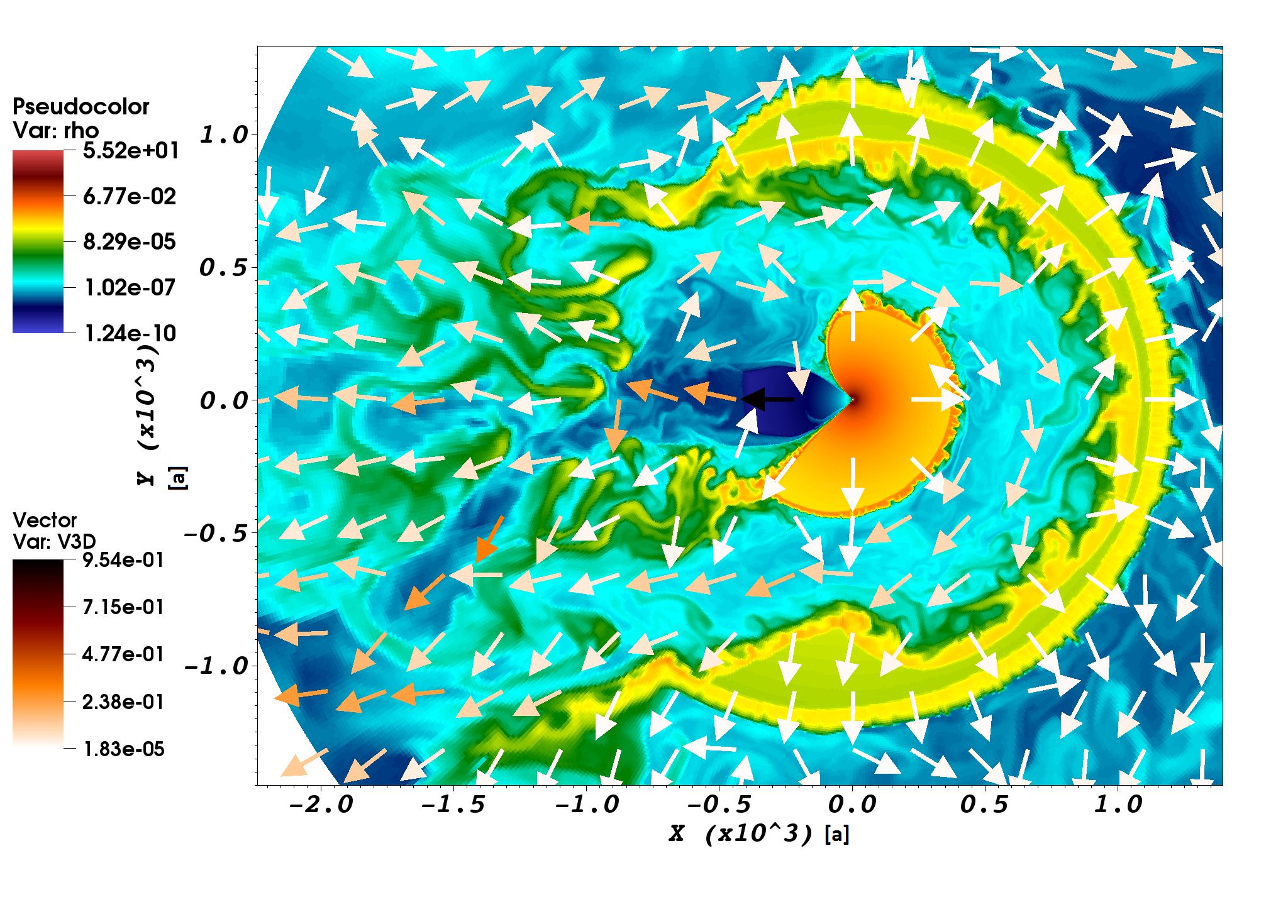}
\includegraphics[width=88mm]{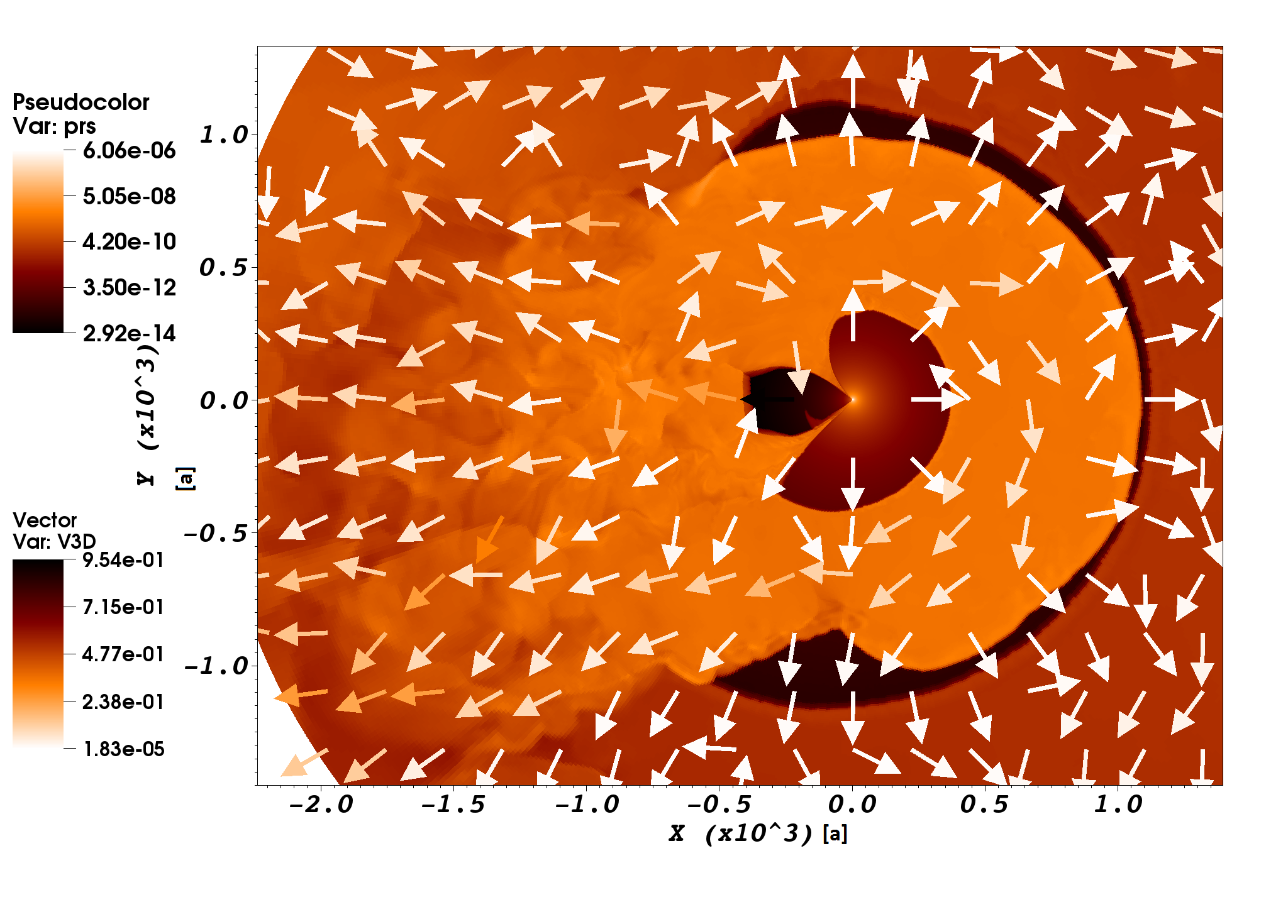}
\includegraphics[width=88mm]{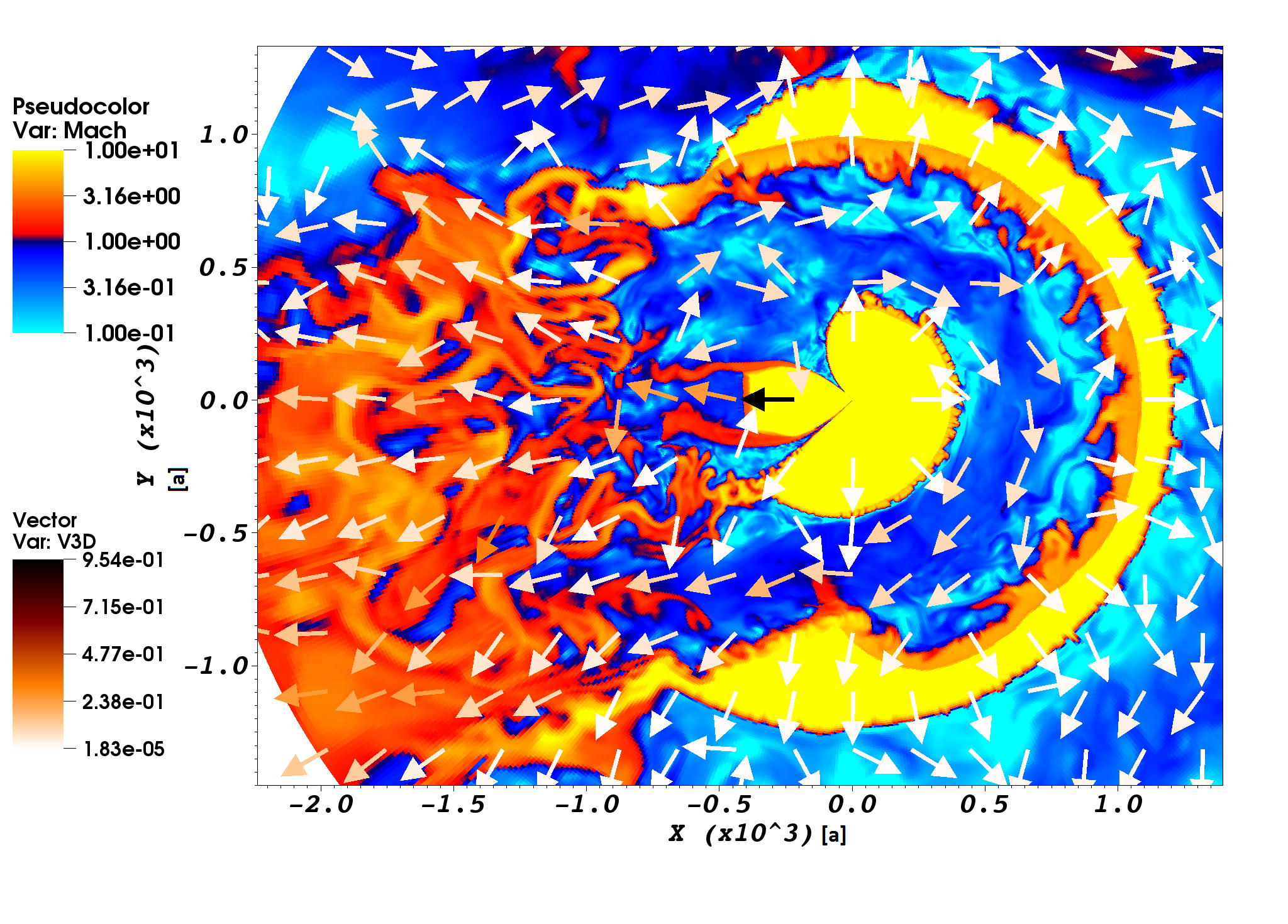}
\vspace{-0.75cm}
\caption{Distribution by colour of density (top panel), pressure (middle panel), and Mach number (bottom panel), and arrows representing the flow motion direction, in the orbital plane (XY, in a units) for \psr{} system 
after 2500 days (680 days after last periastron passage).}
\label{fig:psrrpm}
\end{figure}

\subsection{Numerical setup}

The problem of two colliding winds with orbital motion presents several characteristic length scales: (i) the thickness of the shocked winds right between the two stars; (ii) the orbital separation distance; (iii) the stellar wind speed times the orbital period; (iv) the termination radius in the ISM of the shocked, mixed winds. Technically, it is not possible nowadays to resolve all these scales simultaneously, and in this work we want to focus on the intermediate scales: from the orbital semi-major axis ($a$) size to wind mixing scales. For this, we use a new approach that requires several simplifications to allow for a feasible simulation time while keeping the results realistic almost quantitatively:
\begin{itemize}
 \item We solve the 3D set of equations of relativistic hydrodynamics in spherical coordinates $(r,\theta,\phi)$ but with a very low resolution in the $\theta$-direction. This
allows us to realistically reproduce the wind density profiles and the formation of boundary layers between the winds on the scales of interest.
 \item The pulsar and the stellar wind are assumed to be radially injected, uniformly in the $\phi$-direction, with a constant opening angle $2\alpha$ and $2\pi-2\alpha$ for the pulsar and the stellar wind, respectively\footnote{The results obtained are partially comparable on the smallest scales, and very similar, to those of previous 2D (slab) and full 3D simulations \citep[see][]{bbkp12,bbp15}}.
 \item The injection radius is placed at $2a$ from the center of the orbital ellipse of the binary, avoiding the central, computationally most expensive, region, so a very long time evolution can be followed.
\end{itemize}


The computational domain has a size $r=[2a,2500a]$, $\theta = [\pi/4,3\pi/4]$, and $\phi = [0,2\pi]$, with resolution 
$[N_{r},N_{\theta},N_{\phi}]=[1152,3,768]$. We are modelling interaction scales larger than the Oe-type star decretion disc. 
Since the mass associated to this disc is few orders of magnitude smaller than that of the stellar wind expelled during $t_{\rm orb}$, 
the presence of the stellar disc has been neglected here, and the stellar wind is assumed to have spherical symmetry. 
The two-wind thrust ratio was fixed to $\eta = 0.1$, the stellar wind speed to $v_{w}=0.008$~c, the gas adiabatic index to $\gamma_{\rm ad}=4/3$, 
the pulsar wind Lorentz factor to $\Gamma = 3$, the
orbital period to $t_{\rm orb}=1237$~days, the eccentricity to $e = 0.87$, and masses of the stars to $30 M_{\sun}$ and $1.44 M_{\sun}$ for the normal star and the pulsar, respectively, which together with $t_{\rm orb}$ imply $a\approx 10^{14}$~cm. The periastron and apastron pulsar locations are to the right and to the left from the point $(0,0)$, respectively.

\subsection{Simulation results}

The simulation runs for two full orbital periods, which results in the density map shown in Fig.~\ref{fig:psrrho}. 
As seen in the figure, to the right from the binary there are two semi-annular low-density regions of shocked pulsar 
wind material injected during $t_{\rm pe}$; these two regions are separated by shocked stellar wind. To the left 
from the binary, the unshocked pulsar wind freely propagates radially until is shocked due to orbital bending of 
the shocked winds, after which the shocked pulsar wind keeps moving without much deflection. Not far beyond that 
point, the mixing of the two winds becomes apparent in the form of a fast moving, low-density flow filled with 
slow fragments of stellar wind material\footnote{The toroidal magnetic field component of the stellar wind may 
prevent fragmentation below a certain scale. For this to happen, the magnetic field on the stellar surface should 
be $\gtrsim 300$~G.}. The main difference between the two sides of the interaction structure is that the 
flow is mixed, moving in the radial direction, in the apastron side, whereas in the periastron side the two winds 
are differentiated, with the shocked stellar wind slowly moving outwards, and the shocked pulsar wind 
(confined by the shocked stellar wind) moving in the azimuthal direction.

The typical radial velocity of the flows moving in the apastron (Sa) and the periastron (Sp) side can be computed averaging over $\phi$ and weighting with density in the orbital plane:
\beq
<V_r> = \frac{\int_{\phi_b}^{\phi_e}{\rho v_r d\phi}}{ \int_{\phi_b}^{\phi_e}{\rho d\phi}}\,.
\label{vavr}
\eeq
The sectors Sa and Sp are chosen to illustrate the typical flow velocities in both sides of the binary, apastron and periastron, and correspond to the $\phi$-ranges $[\phi_b,\phi_e]=[2.62,3.66]$ and $[\phi_b,\phi_e]=[5.07,6.12]$, respectively. The angular intervals are equal, and $\phi$ grows counterclockwise from periastron. The results are presented in Fig.~\ref{fig:psrvav}, where a fast outflow is seen in the apastron side, with a speed of $\approx 0.08$~c, much faster than the outflow in sector Sp. Note that the value of $<V_r>$ is dominated by the stellar wind fragments, which density is many orders of magnitude higher than that of the pulsar wind.

Figure~\ref{fig:psrrpcm} shows the radial profiles of the $\phi$-averaged, volume weighted, pressure and Mach number for sectors Sa and Sp. Two slightly different simulation times are shown for Sa to illustrate that, despite the chaotic flow behavior, the profiles are not strongly sensitive to orbital phase in the apastron side.
Whereas for Sp the pulsar and the stellar wind are clearly differentiated, the former presenting constant pressure and low Mach numbers, for Sa there is first a region dominated by the shocked, re-accelerated pulsar wind, followed by a mixed-wind zone with a smooth pressure decrease at $r>500\,a$, meaning that the flow is getting faster and more supersonic. At $r>1700\,a$, the profiles are less reliable as boundary effects may still be relevant.

Figure~\ref{fig:psrrpm} illustrates the importance of wind mixing and acceleration. In particular, in the apastron side the figure shows wind mixing (top) and gradual acceleration due to a gradient of pressure (middle), with the flow becoming supersonic further out (bottom).  


\section{Summary and Discussion}\label{disc}

As proposed in Sect.~\ref{model} and confirmed by our simulations, the highly eccentric orbit in \psr{} leads to a strong anisotropy in the stellar-pulsar wind interaction structure. In the periastron side, intercalated semi-annular layers of shocked pulsar and stellar wind slowly flow outwards, keeping coherent on the simulated scales. Otherwise, on the apastron side, the pulsar and the stellar wind quickly mix, forming a fast and inhomogeneous outflow moving on average in the apastron direction at $\approx 0.08$~c. As this outflow should be rich in shocks and a potential site of particle acceleration, it was probably related to the X-ray-emitting object observed by \cite{phk15}, which found a radial velocity very similar to the value predicted here, and derived an ejection time from the system coinciding with periastron passage. Note that the recurrent ejections of stellar material in the apastron side predicted by us, coinciding with orbital phases around periastron, may have a changing orbit-to-orbit appearance given the strong non-linearity of the dynamical processes involved. Due to the asymmetry of the interaction structure, the flow in the apastron side strongly dominates the energy flux, which should lead to the formation of a one-sided jet-like structure on scales $>1000\,a\approx 10^{17}$~cm. The energetics of such a structure would dominate the interaction with the ISM.

\begin{figure}
\includegraphics[width=90mm]{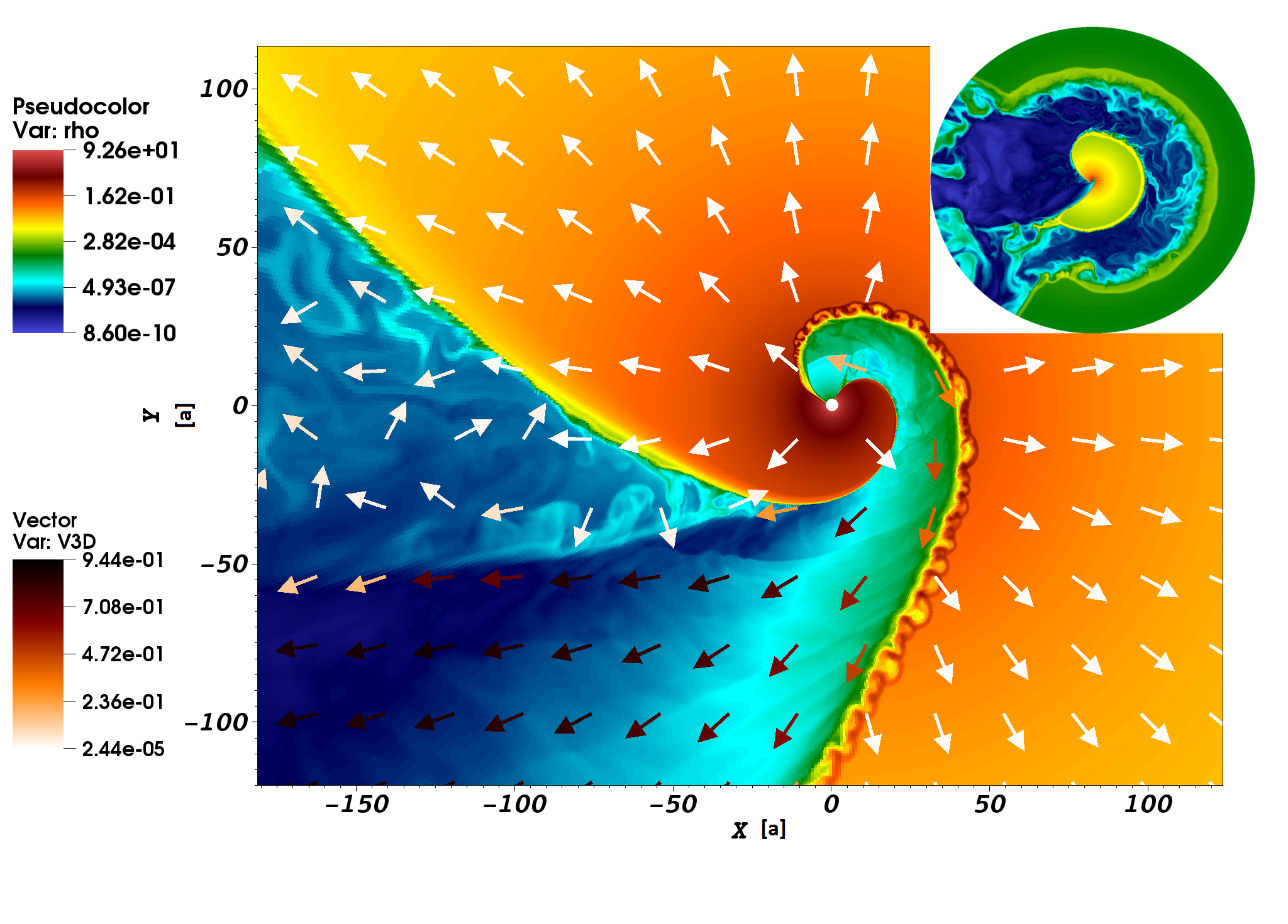}
\vspace{-1cm}
\caption{Zoom in of the density distribution by colour, with arrows representing the flow motion direction, in the orbital plane (XY, in a units) for \psr{} system for $t=1850$~days (27.5 days after last periastron passage). The inset (top, right) shows the same on scales about ten times larger.}
\label{fig:psrrhopp}
\end{figure}

In addition to the X-ray findings, radio observations are also an important tool to study the morphology of the interaction structure in \psr{}. Close to periastron passage, the bending distance of the shocked two-wind structure is within $2a$ and these phases cannot be accurately treated in our approach. For slightly earlier and later phases, however, the bending distance becomes $>2a$ and the geometry of the turning flows is adequately captured. A zoom in on this region, together with the large-scale pattern (see inset), is shown in Fig.~\ref{fig:psrrhopp} for one month after periastron passage. The scales of the turnover of the shocked winds are about $10^{15}$~cm and similar to those of the radio structures found by \cite{mol11}, which suggests that these structures may be actually related to this turnover, although the geometry projected in the sky is unknown, preventing a direct comparison. As seen also in the inset of the figure, even for such an orbital phase, near periastron, a cone formed by fast shocked pulsar wind is still present in the apastron side on larger scales, as suggested in Sect.~\ref{model}.

\section{Acknowledgments}
The calculations were carried out in the CFCA cluster of National Astronomical Observatory of Japan and Greate Wave FX100 Fujitsu cluster in RIKEN. 
We thank Andrea Mignone and the
{\it PLUTO} team for the possibility to use the {\it PLUTO} code and for technical support. 
The visualization of the results performed in the VisIt package \citep{HPV:VisIt}. 
This research has been supported by the Marie Curie Career Integration Grant 321520.
M.B. acknowledges partial  support  by the JSPS (Japan Society for the Promotion of Science):
No.2503786, 25610056, 26287056, 26800159. BMV also acknowledges MEXT (Ministry of Education, Culture, Sports, Science and Technology):
No.26105521 and RFBR grant 12-02-01336-a.
V.B-R. acknowledges support by the Spanish Ministerio de Econom\'{\i}a y Competitividad (MINECO) under grant AYA2013-47447-C3-1-P, and MDM-2014-0369 of ICCUB (Unidad de Excelencia 'Mar\'ia de Maeztu'). 
V.B-R. also acknowledges financial support from MINECO and European Social Funds through a Ram\'on y Cajal fellowship.
\bibliographystyle{mn2e}
\bibliography{text}

\begin{thebibliography}{}

\bibitem[\protect\citeauthoryear{{Bogovalov}, {Khangulyan}, {Koldoba},
  {Ustyugova} \& {Aharonian}}{{Bogovalov} et~al.}{2012}]{bog12}
{Bogovalov} S.~V.,  {Khangulyan} D.,  {Koldoba} A.~V.,  {Ustyugova} G.~V.,
  {Aharonian} F.~A.,  2012, \mnras, 419, 3426

\bibitem[\protect\citeauthoryear{{Bogovalov}, {Khangulyan}, {Koldoba},
  {Ustyugova} \& {Aharonian}}{{Bogovalov} et~al.}{2008}]{bog08}
{Bogovalov} S.~V.,  {Khangulyan} D.~V.,  {Koldoba} A.~V.,  {Ustyugova} G.~V.,
   {Aharonian} F.~A.,  2008, \mnras, 387, 63

\bibitem[\protect\citeauthoryear{{Bosch-Ramon} \& {Barkov}}{{Bosch-Ramon} \&
  {Barkov}}{2011}]{bb11}
{Bosch-Ramon} V.,  {Barkov} M.~V.,  2011, \aap, 535, A20

\bibitem[\protect\citeauthoryear{{Bosch-Ramon}, {Barkov}, {Khangulyan} \&
  {Perucho}}{{Bosch-Ramon} et~al.}{2012}]{bbkp12}
{Bosch-Ramon} V.,  {Barkov} M.~V.,  {Khangulyan} D.,    {Perucho} M.,  2012,
  \aap, 544, A59

\bibitem[\protect\citeauthoryear{{Bosch-Ramon}, {Barkov} \&
  {Perucho}}{{Bosch-Ramon} et~al.}{2015}]{bbp15}
{Bosch-Ramon} V.,  {Barkov} M.~V.,    {Perucho} M.,  2015, \aap, 577, A89

\bibitem[\protect\citeauthoryear{Childs, Brugger, Whitlock, Meredith, Ahern,
  Pugmire, Biagas, Miller, Harrison, Weber, Krishnan, Fogal, Sanderson, Garth,
  Bethel, Camp, R\"{u}bel, Durant, Favre \& Navr\'{a}til}{Childs
  et~al.}{2012}]{HPV:VisIt}
Childs H.,  Brugger E.,  Whitlock B.,  Meredith J.,  Ahern S.,  Pugmire D.,
  Biagas K.,  Miller M.,  Harrison C.,  Weber G.~H.,  Krishnan H.,  Fogal T.,
  Sanderson A.,  Garth C.,  Bethel E.~W.,  Camp D.,  R\"{u}bel O.,  Durant M.,
  Favre J.~M.,    Navr\'{a}til P.,  2012, in , {High Performance
  Visualization--Enabling Extreme-Scale Scientific Insight}.
pp 357--372

\bibitem[\protect\citeauthoryear{{Eichler} \& {Usov}}{{Eichler} \&
  {Usov}}{1993}]{eu93}
{Eichler} D.,  {Usov} V.,  1993, \apj, 402, 271

\bibitem[\protect\citeauthoryear{{Johnston}, {Manchester}, {Lyne}, {Bailes},
  {Kaspi}, {Qiao} \& {D'Amico}}{{Johnston} et~al.}{1992}]{joh92}
{Johnston} S.,  {Manchester} R.~N.,  {Lyne} A.~G.,  {Bailes} M.,  {Kaspi}
  V.~M.,  {Qiao} G.,    {D'Amico} N.,  1992, \apjl, 387, L37

\bibitem[\protect\citeauthoryear{{Mignone} \& {Bodo}}{{Mignone} \&
  {Bodo}}{2005}]{mig05}
{Mignone} A.,  {Bodo} G.,  2005, \mnras, 364, 126

\bibitem[\protect\citeauthoryear{{Mignone}, {Bodo}, {Massaglia}, {Matsakos},
  {Tesileanu}, {Zanni} \& {Ferrari}}{{Mignone} et~al.}{2007}]{mbm07}
{Mignone} A.,  {Bodo} G.,  {Massaglia} S.,  {Matsakos} T.,  {Tesileanu} O.,
  {Zanni} C.,    {Ferrari} A.,  2007, \apjs, 170, 228

\bibitem[\protect\citeauthoryear{{Mold{\'o}n}, {Johnston}, {Rib{\'o}},
  {Paredes} \& {Deller}}{{Mold{\'o}n} et~al.}{2011}]{mol11}
{Mold{\'o}n} J.,  {Johnston} S.,  {Rib{\'o}} M.,  {Paredes} J.~M.,    {Deller}
  A.~T.,  2011, \apjl, 732, L10

\bibitem[\protect\citeauthoryear{{Negueruela}, {Rib{\'o}}, {Herrero},
  {Lorenzo}, {Khangulyan} \& {Aharonian}}{{Negueruela} et~al.}{2011}]{neg11}
{Negueruela} I.,  {Rib{\'o}} M.,  {Herrero} A.,  {Lorenzo} J.,  {Khangulyan}
  D.,    {Aharonian} F.~A.,  2011, \apjl, 732, L11

\bibitem[\protect\citeauthoryear{{Pavlov}, {Hare}, {Kargaltsev}, {Rangelov} \&
  {Durant}}{{Pavlov} et~al.}{2015}]{phk15}
{Pavlov} G.~G.,  {Hare} J.,  {Kargaltsev} O.,  {Rangelov} B.,    {Durant} M.,
  2015, \apj, 806, 192

\end{thebibliography}
\end{document}